# Molecular beam brightening by shock-wave suppression

Yair Segev, Natan Bibelnik, Nitzan Akerman, Yuval Shagam, Alon Luski, Michael Karpov, Julia Narevicius and Edvardas Narevicius*

Department of Chemical Physics, Weizmann Institute of Science, Rehovot 7610001, Israel

**Supersonic beams are a prevalent source of cold molecules utilized in the study of chemical reactions, atom interferometry, gas-surface interactions, precision spectroscopy, molecular cooling and more. The triumph of this method emanates from the high densities produced in relation to other methods, however beam density remains fundamentally limited by interference with shock waves reflected from collimating surfaces. Here we show experimentally that this shock interaction can be reduced or even eliminated by cryo-cooling the interacting surface. An increase in beam density of nearly an order of magnitude was measured at the lowest surface temperature, with no further fundamental limitation reached. Visualization of the shock waves by plasma discharge and reproduction with direct simulation Monte Carlo calculations both indicate that the suppression of the shock structure is partially caused by lowering the momentum flux of reflected particles, and significantly enhanced by the adsorption of particles to the surface. We observe that the scaling of beam density with source pressure is recovered, paving the way to order of magnitude brighter cold molecular beams.**

## Introduction

Atomic and molecular beams have enabled groundbreaking experiments in physics and chemistry for almost a century. Pioneering applications of these methods such as the Stern-Gerlach experiment[1], and later the maser[2], employed effusive beams. These beams were inherently limited to low densities and relatively high temperatures as particles would leave the source in free-molecular flow. Supersonic nozzles[3] introduced dramatically denser ("brighter") beams of significantly colder particles, leading to major advances in various fields such as reaction studies[4], nanomaterials[5], and atom interferometry[6]. More recently supersonic beams have been applied as sources for decelerators[7,8] used for novel methods of molecular cooling[9], in the study of quantum phenomena in cold collisions[10–12], and for molecular interferometry[13].

The success of supersonic beams stemmed largely from the high densities they offered; however, continuous flow (CW) experiments often remained limited by maintainable background pressure, as they required high pumping speeds[14]. Advances in pulsed nozzles have effectively removed these restrictions[15], enabling the use of source pressures up to three orders of magnitude higher than with a CW beam. However, beam brightness seems only to have reached a new limit, induced by interference of the beam with necessary collimating diaphragms known as skimmers[16]. Sufficiently high densities lead to "clogging", an interaction of the flow through the skimmer with particles reflected from the skimmer walls that manifests as shock waves[17] and limits the beam flux[18]. Thus, common practice with pulsed beams remains to this day to limit the beam brightness by either restricting source pressure, distancing the nozzle from the skimmer, or both[16].

* edn@weizmann.ac.il

Here we demonstrate that the shock wave interaction with the skimmer and the associated clogging can be significantly reduced or even eliminated by cryo-cooling the skimmer. We trace the elimination of clogging to two effects, each of which lowers the momentum flux reflected from the surface. Cooling the surface reduces the thermal velocity of reflected particles and decreases their mean free path into the undisturbed beam. Still lower temperatures lead to adsorption of the impacting particles to the skimmer. We observe that the suppression of shock waves by reduction of reflected momentum is possible since the shock propagation in a rarefied medium is governed by direct collisions between the beam and reflected particles. This mechanism is in stark contrast with the well-studied propagation of shocks through a supersonic fluid in the continuum regime. We measure a resulting increase in beam flux of a factor of 8, with no further fundamental limitation reached in our experiment.

**Formation of shocks in supersonic beams**

Supersonic beams are generated by releasing gas through a nozzle into vacuum. Flow emanating from the stagnant source into the low-pressure volume expands due to collisions between flow particles, leading to narrowing of the velocity spread from the source's thermal distribution to a low-temperature, high velocity beam. At any point in the early expansion region, the flow conditions – velocity magnitude, density and temperature – are connected through the isentropic relations to the local Mach number and the stagnation conditions[19]. The free-jet shock structures present in most continuous beams vanish at the sufficiently low background pressures commonly obtained with low duty-cycle pulsed valves[14], and the isentropic expansion continues until the distance between collisions as measured in the lab frame is non-negligible. At this point, typically occurring at distances of hundreds of nozzle throat diameters, the flow gradually "freezes"[20] – the temperature gradient eases while the density continues to decrease as the square of the distance from the nozzle. For most applications it is advantageous to conduct experiments as early as possible in this rarefied region, making use of the maximum available density. However, supersonic expansion creates a broad plume of gas, so the axial part of the beam must first be "skimmed" from the plume in order to produce a collimated jet in the test chamber.

Typical skimmers are straight or flared[21] cones, or 2D slits[22], with the common intention of minimizing the interaction of the centerline region of the beam with particles reflected from either the inner surface into the skimmer or from the outer surface into the oncoming plume. If the beam is sufficiently bright, both reflection directions will manifest as shock waves[17] - abrupt disruptions in the supersonic flow that introduce entropy and heat from the walls. In the interior of the skimmer the shock waves from different points on the lip may combine to clog the skimmer, slowing and heating the beam before a second expansion process follows downstream. The resulting jet will have different terminal properties[23], as the stagnation temperature of the beam will be changed by the energy introduced from the surfaces. Common practice is to avoid this "skimmer interference" by restricting the density at the skimmer entrance, either by increasing the distance between the nozzle and the skimmer or by decreasing the source pressure. In this manner skimmer interference is often the limiting factor in creating bright cold beams.

In order to reduce the effect of skimmer interference without restricting the density, we propose to cool the skimmer wall. Use of a cooled skimmer was previously tested with a CW beam at the Arnold Engineering Development Complex[24–27], however, reaching sufficient source pressures for skimmer interference with a continuous source required enormous pumping speeds in excess of 500,000 *L/s*, and quickly led to filling of the skimmer orifice with frozen particles[24].

**Results and discussion**

The experimental setup is illustrated in Figure 1 (a). The beam source is an Even-Lavie valve[28,29] with a nozzle diameter of 0.2*mm*, attached to a gas cylinder of variable stagnation pressure $P_0$. The skimmer is attached to a 10K cryostat and held at a temperature of $T_w$, monitored with a diode and maintained with a heating resistor. In some of the experiments the valve is attached to a second cryostat, to vary the stagnation temperature $T_0$, while in others it is attached to a manipulator, controlling the nozzle to skimmer distance $r_{ns}$. The skimmer (Figure 1, b), made of Oxygen-free high thermal conductivity copper, is a cone with a tapered wall - thick near the base for sufficient heat conduction, and sharpening towards the tip. The entry diameter of this skimmer is $D_s$=4*mm*.

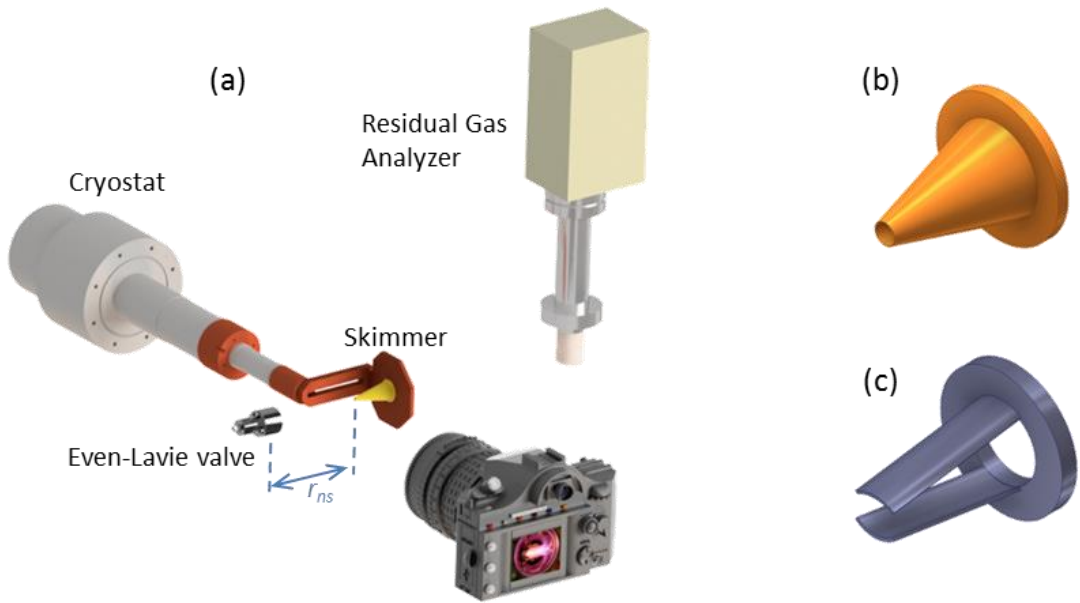

**Figure 1:** Experimental setup for measuring the effect of skimmer temperature on the transmitted beam. The layout (a) includes a pulsed valve, a cryo-cooled skimmer, and a residual gas analyzer for beam density measurement; vacuum chambers are omitted from the illustration for clarity. A camera images the plasma glow induced by a discharge pulse, revealing shock waves in the density field. A conical copper skimmer (b) is used for comparative density measurements, while a split aluminum skimmer (c) is used for visualization of the internal shock structure.

A residual gas analyzer (RGA) is used to measure the density of the beam 215*mm* downstream of the skimmer entrance. We define a case as "unclogged" if an increase in source pressure by a factor of *k* leads to a similar increase in peak density, and "clogged" if the increase is significantly smaller than *k*. Measurements of clogged beams at varying $T_w$, shown in Figure 2, reveal a sensitivity of the density to the skimmer temperature for different gases from constant source conditions. At high surface temperatures the measured beams exhibit a sharp peak, followed by a dip and a long "tail" lasting hundreds of milliseconds. As the temperature of the skimmer is lowered the first peak initially increases while the tail gradually diminishes, with increasing sensitivity at lower temperatures. Below a certain temperature the density profile stops changing. Concurrently, we observe that the peak density becomes sensitive to source pressure, so the skimmer is now effectively unclogged. This temperature, termed the "unclogging temperature" $T_u$, varies with species – for Krypton it is approximately 60*K*, for Argon 40*K*, and for Neon 20*K*.

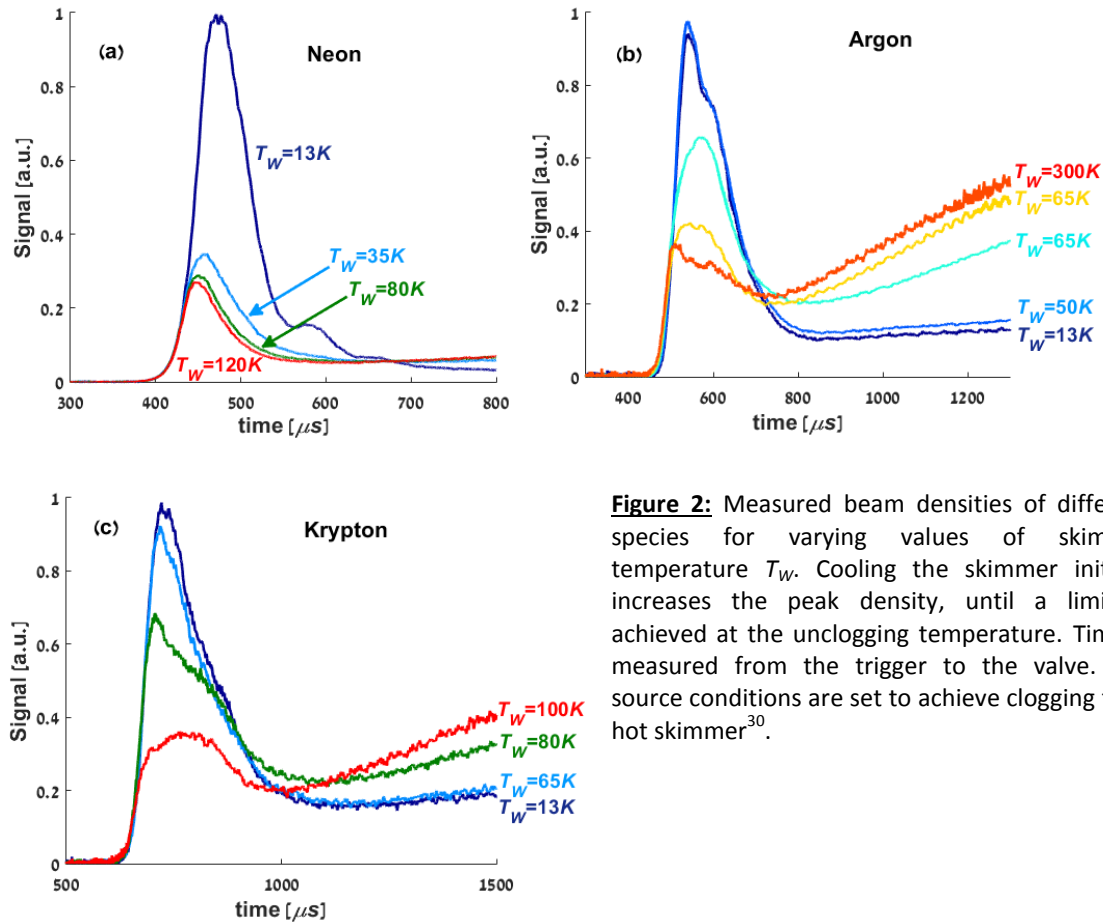

**Figure 2:** Measured beam densities of different species for varying values of skimmer temperature $T_W$. Cooling the skimmer initially increases the peak density, until a limit is achieved at the unclogging temperature. Time is measured from the trigger to the valve. The source conditions are set to achieve clogging for a hot skimmer[30].

The variation in beam density corresponds to a change in the flow structure through and around the skimmer. To view the flow phenomena responsible for this, we add a pulsed-discharge grid in front of the valve. The plasma glow from the denser portions of the flow reveals the oblique shock waves that envelop the skimmer (Figure 3). At high temperatures (Figure 3, a-b), the angle between the surface and the shock is far larger than the corresponding shock angle in flows of similar Mach number but lower Knudsen number. This indicates that the disturbance to the beam propagates directly by collisions with reflected particles, even at distances comparable with the skimmer dimensions. This effect is a result of the large mean free path of particles reflected from the surface into a rarefied flow, in contrast with the typical shock waves of the continuum regime, which propagate at the thermal velocity of the medium. At lower surface temperatures the shock angle significantly decreases (Figure 3, d-e), approaching the characteristic acute angles expected at high Mach numbers. Near or below $T_u$ the shock is almost parallel to the skimmer, and may even vanish entirely (Figure 3, g-h). In this regime condensation of frozen beam particles gradually appears on the skimmer over time.

The variation of the external shock angle with skimmer temperature, Figure 4, shows a general trend similar to that of the density profile, with increasing sensitivity as the temperature decreases to $T_u$. This correlation derives from a near mirroring of the external shock that occurs in the flow within the skimmer. Utilizing a "split skimmer" (Figure 1, c), we observe oblique shock waves emanating from the lip and intersecting at the centerline (Figure 3, c), a flow structure identified with skimmer clogging[17]. As we lower the skimmer temperature the high density region behind the shock fronts becomes less intense and more elongated, and the shocks recline (Figure 3, f). This structure eventually vanishes below $T_w$ (Figure 3, i).

Further insight into the mechanism of skimmer interference and its dependence on temperature emerges from numerical simulation of the flow using the direct simulation Monte Carlo (DSMC) method. We employ the DS2V code[31] with unsteady sampling, species-dependent parameters from ref. (32), and beam parameters estimated from analytical and semi-empirical expressions for different stagnation conditions[19,33]. Reflections from the skimmer surface are defined as fully diffuse at temperature $T_w$, initially assuming no adsorption.

The calculated flow exhibits shock waves for modified Knudsen numbers[34] of approximately unity and less. Figure 5 presents such cases for a Krypton beam impacting hot (a) and cold (b) conical skimmers. With decreasing skimmer temperature the shocks waves recline, and the internal shock structure elongates, qualitatively reproducing the experimentally observed trend. In the simulation, the mechanism behind this trend can only derive from the change in the thermal velocity of reflected particles, with lower temperature leading to lower mean free path of these particles and more acute shock angles.

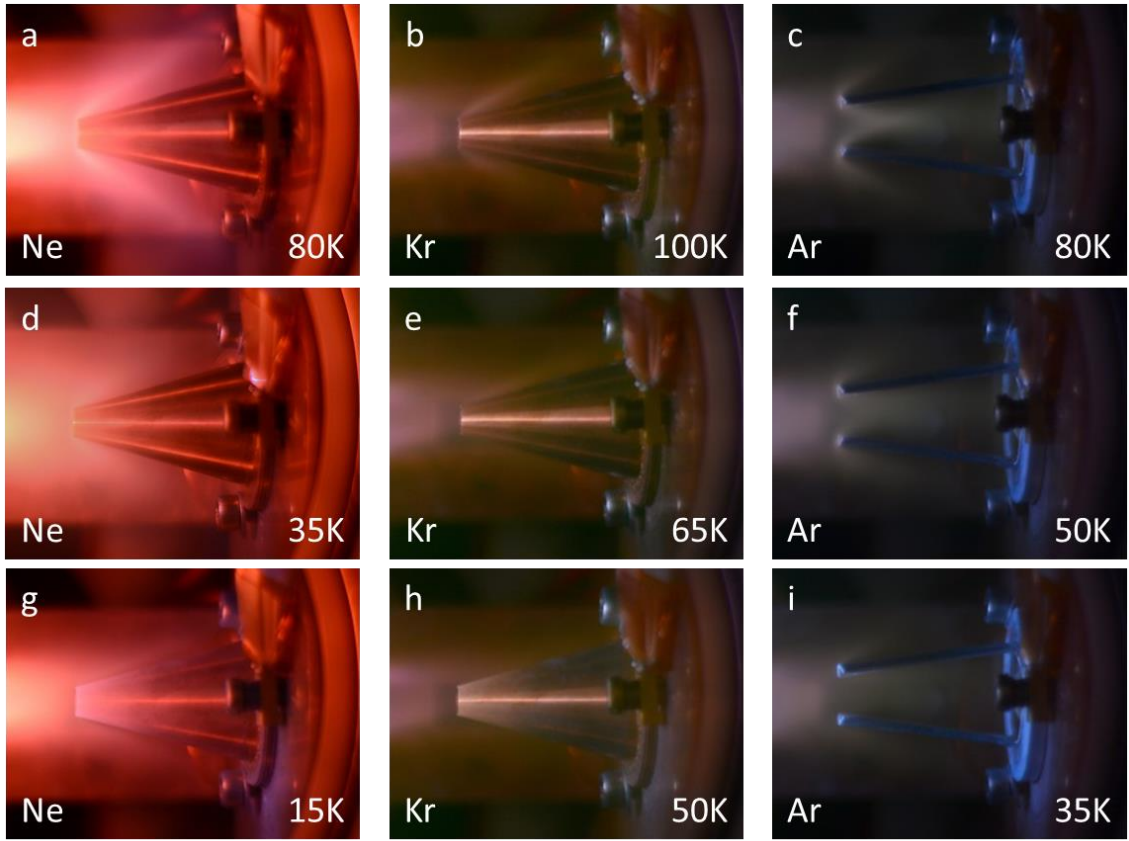

**Figure 3:** Visualization of the density field using a pulsed discharge. Shock waves enveloping the skimmer recline and eventually vanish as the skimmer temperature is lowered. The internal shock structure exhibits similar phenomena. For each case the species is indicated in the lower left corner, and skimmer temperature in the lower right. The source conditions are set to achieve clogging at high temperature[35].

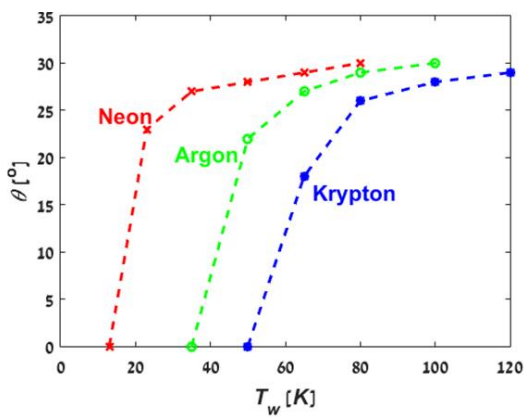 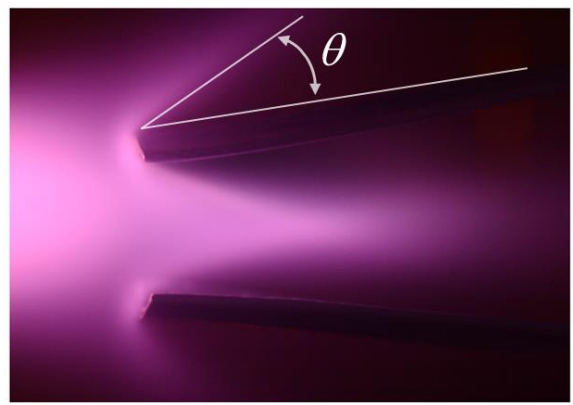

**Figure 4:** External shock wave angle as a function of skimmer temperature. The shock angle, defined on the right over an image with a Helium beam, initially decreases gradually with the temperature. Near the unclogging temperature the sensitivity increases, until the shock becomes parallel to the wall and vanishes. The source conditions are set to achieve clogging at high temperature[35].

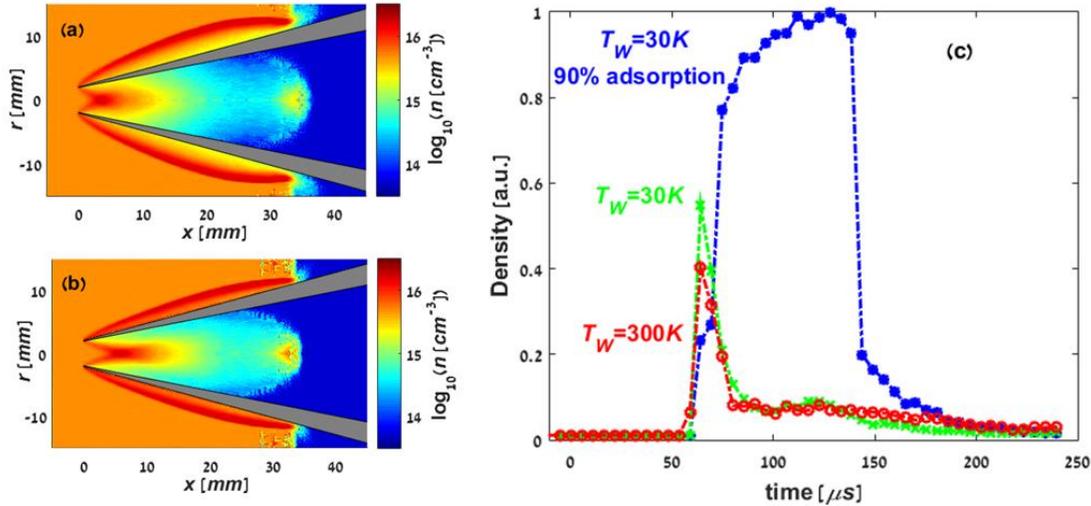

**Figure 5:** Direct simulation Monte Carlo calculations of a Krypton beam impacting a conical skimmer. The density field is presented in logarithmic scale for cases of a "hot" ($T_w$=300K, (a)) and a "cold" ($T_w$=30K, (b)) skimmer. The transmitted beam consists of a dense "bullet" followed by a sparse "cloud" corresponding to the peaks and "tails" in experimentally measured beams. Changes in the shock structures with skimmer temperature qualitatively match the experimental observations above the unclogging temperature. The centerline density 30$mm$ downstream of the skimmer entrance (c) also exhibits a beam intensification trend at lower temperatures, but only the addition of significant adsorption to the surfaces enables complete unclogging. Supplementary parameters are defined in ref. (36).

The transmitted beam in the simulated skimmers presents a relatively dense "bullet" followed by a hot, sparser "cloud" of expanding gas. The bullet corresponds to the portion of the beam that passes the skimmer entrance before the shock structure has completely formed, while the cloud corresponds to the re-expansion of a jet behind a strong shock. The intensity and length of the "bullet" both increase as $T_w$ is decreased, whereas the total flux emitted later from the "cloud" decreases. Thus, for a pulsed beam sampled in time at a constant point downstream of the shock structure, as in Figure 5 (c), the density qualitatively matches the peaks and "tails" measured experimentally and presented in Figure 2.

The duration of the transmitted portion of the beam peak is limited by the shock formation time, $t_s \approx D_s/2v_t$, where $v_t$ is the thermal velocity of the gas arriving from a skimmer lip held at temperature $T_w$ to the centerline. This expression holds as long as the mean free path of reflected particles and the skimmer radius are of the same order of magnitude. Indeed the measured rise times of the transmitted peaks roughly match the corresponding values of $t_s$, decreasing with lower $T_w$ and lighter species, as evident in Figure 2. The skimmer transmits higher peak densities when $t_s$ is a sufficiently large fraction of the full pulse rise time. Since $t_s$ scales with the square root of molecular mass, the shock formation times at the lowest temperatures available in the current experiment are sufficiently long to observe the onset of unclogging in beams of Neon, Argon, and Krypton, but unclogging a Helium beam requires even lower temperatures. Note that an additional consequence of a finite clogging time is that when skimmer clogging occurs only the faster portion of a pulsed beam is transmitted without interference. This effect is detrimental for applications where a lower mean beam velocity is beneficial, such as decelerators.

The mechanism of decreased thermal velocity alone fails to predict the existence of a complete unclogging temperature, since the transmitted pulse duration is far longer than $t_s$ at the values of $T_u$ relevant for each species. Furthermore, the rapid reduction in shock angle as $T_w$ approaches $T_u$, followed by vanishing of the shock, significantly exceeds the numerically predicted effect. It appears that the onset of significant adsorption enhances the reduction in reflected momentum, and explains the gradual appearance of condensation on the skimmer at these temperatures. Adding sufficient adsorption fractions to the surfaces in the DSMC calculations also reproduces unclogged skimmers (Figure 5, c). We conclude that at sufficiently low temperatures skimmer clogging is relieved by condensation of many or most of the impacting particles.

In order to demonstrate that beam transmission with a cold skimmer is no longer limited by density, we present the peak measured density with increasing source pressure (Figure 6, a) and with decreasing nozzle to skimmer distance (Figure 6, b). At either large distances or low source pressures, a room-temperature skimmer and a skimmer cooled below $T_u$ identically transmit the beam. However, larger densities lead to clogging of the "hot" skimmer, causing a leveling off of the density. In contrast, the peak transmitted density with a cold skimmer continues rising without any observed limit. Our measurements have found an increase of a factor of 8 with Neon from a maximum source pressure of 85$atm$ and a minimum $r_{ns}$ of 28$mm$. We expect that significantly higher densities can yet be achieved with higher source pressures.

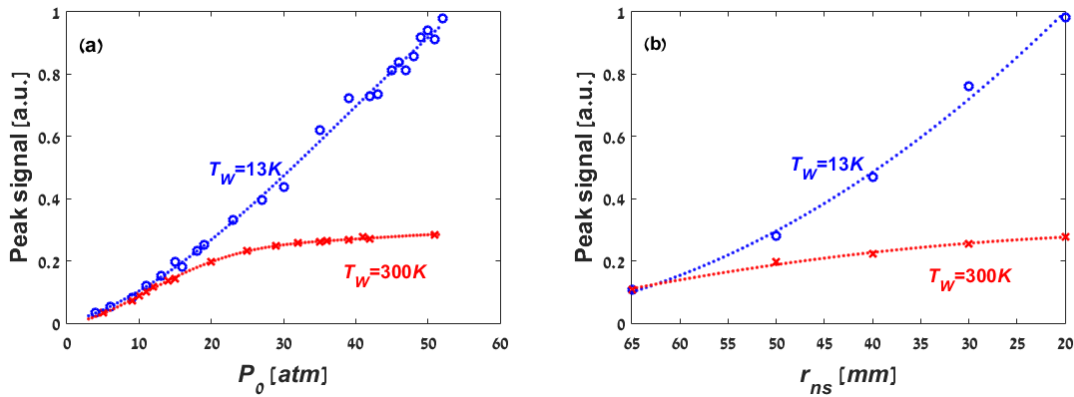

**Figure 6:** Peak measured density for a Neon beam transmitted through "hot" ($T_w$=300$K$) and "cold" ($T_w$=13$K$) conical skimmers with varying beam density. The density at the skimmer entrance is controlled either by varying the source pressure (a) at a constant nozzle-skimmer distance ($r_{ns}$=45$mm$), or by varying this distance (b) at a constant stagnation pressure ($P_0$=40$atm$). Increasing the density by either method eventually clogs the hot skimmer, whereas the cold skimmer transmits a continuously increasing peak density with no observed limit. The source temperature is 300$K$.

## Summary and outlook

We have demonstrated the successful suppression of the shock waves that frequently interfere with molecular beam experiments. By cryo-cooling the collimating surfaces, the flux of reflected momentum causing this interference was reduced, resulting in the revival of density scaling with source pressure. An increase in density by a factor of 8 was measured, with no new fundamental limit reached. The removal of this shock wave interference lifts the long standing limitation on the brightness of molecular beams.

Boosting the beam brightness by an order of magnitude is beneficial for a wide variety of experiments. Higher densities can provide increased sensitivity in atom and molecular interferometry. In the case of bimolecular collisions, the collision rates will in fact be increased by two orders of magnitude, assisting in the study of slow processes that proceed by tunneling. Finally, attaining higher beam density may bring within reach new and general molecular cooling methods based on collisions, such as sympathetic or evaporative cooling.

## Acknowledgements

The authors acknowledge financial support from the European Commission through ERC grant EU-FP7-ERC-CoG 1485, and from the Israel Science Foundation through grant 1810/13.